\documentclass[journal]{IEEEtran}
\usepackage[T1]{fontenc}
\usepackage{amsmath}
\usepackage{algorithmic}
\usepackage{algorithm}
\usepackage{array}
\usepackage[caption=false,font=normalsize,labelfont=sf,textfont=sf]{subfig}
\usepackage{textcomp}
\usepackage{stfloats}
\usepackage{url}
\usepackage{verbatim}
\usepackage{graphicx}
\usepackage{xcolor}
\usepackage{cite}
\usepackage{booktabs}
\usepackage{multirow}
\usepackage{tikz}
\usetikzlibrary{shapes,arrows,positioning,fit}
\begin{document}

\title{Bridging the Cognitive Gap: A Unified Memory Paradigm for 6G Agentic AI-RAN}

\author{
    Xijun~Wang,
    Zhaoyang Liu,
    Chenyuan~Feng,
    Xiang~Chen,
    Howard~H.~Yang,
    and Tony~Q.~S.~Quek
%\thanks{This work has been submitted to the IEEE for possible publication. Copyright may be transferred without notice, after which this version may no longer be accessible.}
%\thanks{This work was supported in part by the National Key Research and Development Program of China under Grant 2022YFB2902004 and in part by the National Natural Science Foundation of China under Grants 62271513 and 62301328. (Corresponding authors: Xijun Wang; Xiaoyun Wang.)}
\thanks{Xijun~Wang, Zhaoyang Liu, and Xiang~Chen are with Sun Yat-sen University, China; Howard~H.~Yang is with Zhejiang University, China; Chenyuan~Feng is with University of Exeter, U.K.; 
Tony~Q.~S.~Quek is with Singapore University of Technology and Design, Singapore, and also with Purple Mountain Laboratories, Nanjing, China.}
\vspace{-0.0em} }

\maketitle

\begin{abstract}
As 6G evolves, the radio access network must transcend traditional automation to embrace agentic AI capable of perception, reasoning, and evolution. A fundamental cognitive gap persists in current disaggregated architectures, where interfaces force the physical layer to compress high-dimensional states into low-dimensional metrics, trapping reasoning agents behind a semantic bottleneck. We propose a unified memory paradigm that dissolves the boundaries between sensing and reasoning by mapping biological memory hierarchies onto heterogeneous computing fabrics. Enabled by emerging coherent interconnects, this approach creates a cognitive continuum where real-time reflexes, near-real-time reasoning, and long-term evolution share state across memory tiers, bridging the gap between real-time responsiveness and long-horizon context for agentic 6G networks.
\end{abstract}

\begin{IEEEkeywords}
Agentic AI, AI-RAN, Unified Memory, Heterogeneous Computing, 6G
\end{IEEEkeywords}

\section{Introduction}
\label{sec:introduction}

6G is transforming the Radio Access Network (RAN) from passive transmission infrastructure into an intelligent entity capable of perception, reasoning, and autonomous action. This transformation is driven by billions of connected devices, stringent latency requirements, and the integration of sensing and communications, collectively challenging traditional rule-based automation. Accordingly, standardization bodies are exploring Artificial Intelligence (AI) and Machine Learning (ML) integration into the air interface \cite{ref2}.

The integration of AI into RAN is discussed under the broader term AI-RAN, which distinguishes AI-for-RAN, AI-on-RAN, and AI-and-RAN \cite{airan_kundu2026,airan_khan2024}. Among these, AI-for-RAN focuses on improving RAN operation, with agentic AI offering a path toward greater network autonomy \cite{new1}. An AI agent differs from static control scripts or reactive ML models in its ability to perceive physical reality, construct a world model, and act autonomously toward goals. Recent demonstrations include LLM-based agents for autonomous remediation \cite{new2} and foundation model-based frameworks supporting hierarchical multi-agent orchestration across network layers \cite{chen2025}. These advances motivate the study of agentic AI-enabled RAN functions.

However, current agentic AI-enabled RAN functions suffer from temporal amnesia and spatial blindness. First, agents lack stateful memory. They operate in the immediate present and cannot retain contextual information across decision cycles. Without a shared mechanism for retaining and retrieving historical context, they may fail to recognize recurring patterns and therefore rely more heavily on current observations. Second, current architectures impose a semantic bottleneck that blinds reasoning agents to physical reality. In the disaggregated RAN setting, the RAN Intelligent Controller (RIC) receives physical-layer information through logical interfaces and service models. These abstractions determine which physical-layer state is exposed and how it is represented. Consequently, rich, multidimensional channel information may reach the RIC only through a few scalar indicators. The RIC therefore observes aggregated symptoms rather than their underlying causes.

Both limitations stem from the absence of a unified memory architecture in the disaggregated RAN setting, where relevant state may not be shared consistently across space and time. In this article, we focus on AI-for-RAN, specifically on agentic AI-enabled RAN functions, and present a memory-centric architecture. We map biological memory hierarchies, including sensory, working, and long-term memory, onto heterogeneous computing fabrics comprising field-programmable gate arrays (FPGAs), graphics processing units (GPUs), and central processing units (CPUs), enabled by coherent interconnects such as Compute Express Link (CXL). This approach supports selected cross-tier state sharing and preserves the spatial structure and temporal context required for cause-aware decision-making. Leveraging this fabric, we construct a multi-scale cognitive architecture spanning real-time reflexes, near-real-time reasoning, and long-term evolution, where upper tiers observe lower-tier states and propagate learned policies downward through shared memory, yielding an integrated cognitive hierarchy for RAN intelligence.

%The rest of this paper proceeds as follows. Section II analyzes the cognitive gap in current AI-RAN architectures. Section III presents the hierarchical multi-scale cognitive framework. Section IV details the unified memory fabric and its physical realization. Section V presents a case study. Section VI discusses open issues, and Section VII concludes the paper.

\begin{figure*}[!t]
\centering
\includegraphics[width=0.71\textwidth]{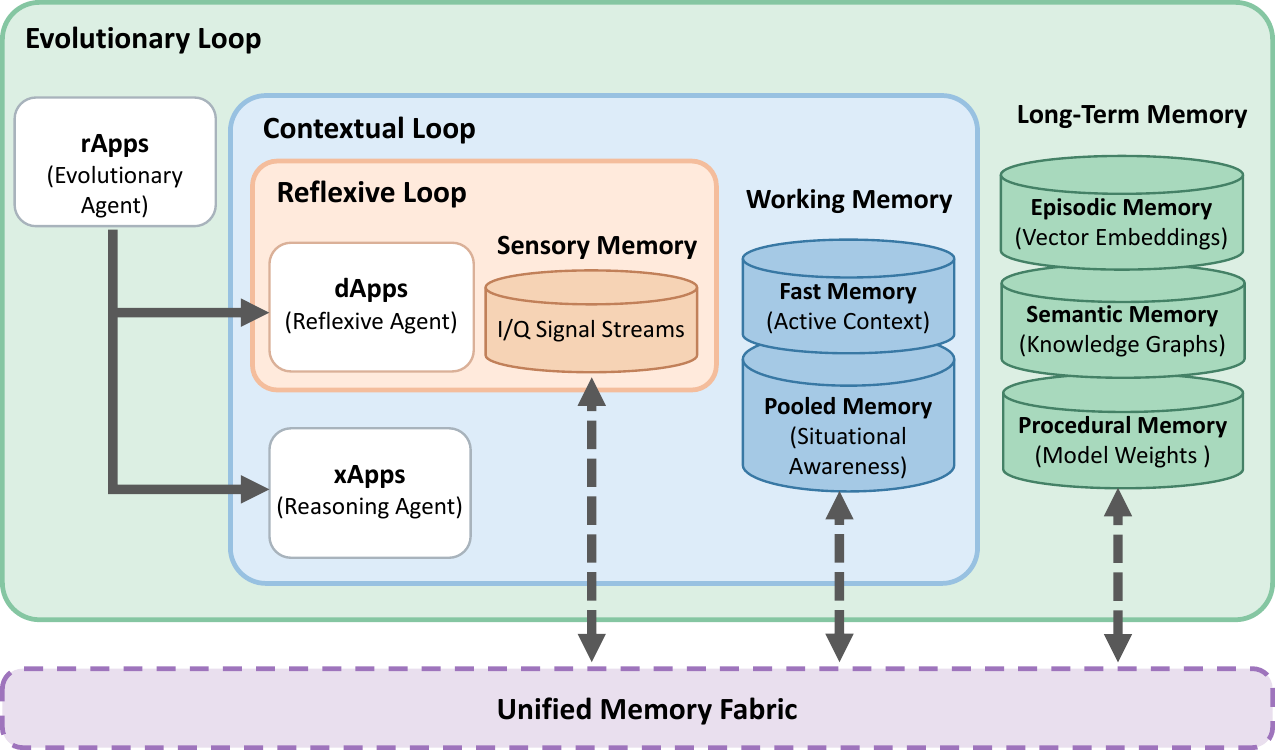}
\caption{Cognitive framework for agentic AI-enabled RAN functions with three nested cognitive loops. The reflexive loop (innermost) handles real-time physical-layer processing with sensory memory. The contextual loop (middle) performs near-real-time planning with working memory. The evolutionary loop (outermost) drives non-real-time learning with persistent storage. All three tiers share state through a unified memory fabric.}
\label{fig:architecture}
\end{figure*}

\section{Cognitive Gaps in Agentic AI-Enabled RAN Functions}
\label{sec:cognitive-gap}

Consider an agent responsible for radio-resource adaptation at a large stadium where similar events recur. To act intelligently, the agent needs two capabilities. It must observe the raw channel state to distinguish co-channel interference from multipath fading, which require different remediation. It must also recall historical patterns, such as traffic surges during past matches, to enable proactive resource reservation. However, current architectures support neither capability. The agent, residing in the RIC, receives only average Signal-to-Interference-plus-Noise Ratio (SINR) through the E2 interface and loses the spatial structure of the channel. It also lacks persistent memory of prior episodes, forcing it to rediscover the same patterns from scratch. %This example illustrates how spatial blindness and temporal amnesia arise in practice.

For an agentic AI-enabled RAN function, spatial blindness arises when physical-layer observations are exposed through interface-defined abstractions. The Open Radio Access Network (O-RAN) Alliance provides openness through disaggregated interfaces such as E2, A1, and O1 \cite{ref6}, all following the principle of information hiding. Under this principle, service-model abstractions expose high-dimensional states, such as beam matrices and channel tensors, to the RIC only through low-dimensional metrics like average SINR or Channel Quality Indicator (CQI) \cite{new4}. Although the E2 Service Model (E2SM) framework is extensible in principle, each new use case requires a dedicated service model before its physical-layer state can be exposed, a process that cannot keep pace with the growing diversity of AI-driven applications. Hence, interface evolution alone is unlikely to close the spatial gap.

This interface-level gap is compounded by a hardware-level gap. In disaggregated deployments, sensing logic on FPGAs or accelerators may operate in a memory domain separate from host CPUs or GPUs running reasoning functions, and cross-domain observation therefore depends on how state is moved or shared. Industry efforts have improved computational efficiency within individual domains. NVIDIA Aerial RAN CoLab (ARC) demonstrates GPU-accelerated Layer 1/Layer 2 (L1/L2) processing with shared memory between the physical layer (PHY) and medium access control (MAC) layers \cite{nvidia_arc}, Intel FlexRAN offloads selected L1 functions to integrated hardware accelerators \cite{intel_flexran}, and O-RAN Working Group 6 has specified cloud-native deployment models for disaggregated RAN \cite{oran_wg6}. These efforts improve within-domain processing but do not provide a common state-sharing path between accelerator and host domains. One possible path forward is CXL, which enables cache-coherent memory sharing across heterogeneous compute units \cite{ref9}. In this setting, its architectural role is to provide coherent load/store access to selected cross-tier state across compatible heterogeneous devices, reducing reliance on explicit message or copy handoffs. Recent work has shown how CXL-attached memory can extend GPU capacity for large-context AI inference \cite{new7}, yet its role in RAN architectures remains largely unexplored.

The same function exhibits temporal amnesia when it cannot retain and retrieve relevant operating context across decision cycles in a disaggregated RAN setting. Wireless environments exhibit structured recurrence, where traffic surges follow event schedules, interference patterns recur with device mobility, and beam configurations that succeeded in one episode often apply to similar future episodes. An AI agent with access to this operational history could anticipate problems rather than merely react to them. Without such retained context, each decision epoch is effectively treated as independent, with the agent discarding experience once the control loop closes. Memory-augmented approaches such as RAN Cortex \cite{new3} represent a step forward by introducing vector-based retrieval engines that allow near-real-time RIC applications (xApps) to query semantically similar historical scenarios during inference. However, because RAN Cortex operates at the xApp level, the agent remembers only high-level symptoms such as throughput degradation rather than underlying physical causes such as beam misalignment in a specific sector. Without physical-layer context, two past episodes may appear identical at the metric level yet arise from entirely different root causes.

\section{Hierarchical Multi-Scale Cognitive Framework}
\label{sec:cognitive-architecture}

In this section, we propose a nested, multi-scale hierarchy that addresses both spatial blindness and temporal amnesia. As shown in Fig.~\ref{fig:architecture}, the system is organized into three cognitive loops interconnected through a unified memory fabric that preserves spatial structure and temporal context. The reflexive, contextual, and evolutionary loops operate at the real-time, near-real-time, and non-real-time horizons of O-RAN, respectively. Through this unified memory fabric, the architecture can retain relevant context and coordinate information exchange across these horizons.

\subsection{Reflexive Loop}
\label{subsec:reflexive-loop}

At the architectural core lies the reflexive loop, responsible for real-time processing. The reflexive loop combines hard reflex implementations for streaming data with soft reflex implementations for more complex real-time tasks. One candidate realization of these functions is a distributed application (dApp), a lightweight and programmable RAN-side software component that accesses fine-grained local radio state and supports bounded-latency real-time functions \cite{dapp_lacava2025}. Unlike the reasoning agents in higher loops, dApps react directly to current radio measurements. During real-time execution, they run pre-trained neural-network models with fixed parameters to produce control actions rather than performing deliberative, long-horizon reasoning. The speed required for this reactive processing is sustained by sensory memory, which serves as a transient buffer for raw, high-fidelity environmental snapshots such as I/Q signal streams and instantaneous beam-level signal-to-noise ratio (SNR) maps. Through the unified memory fabric, dApps expose selected state to the contextual loop, allowing xApps to access relevant sensory information without relying solely on message-level summaries.

\subsection{Contextual Loop}
\label{subsec:contextual-loop}

Enveloping the reflexive core is the contextual loop, operating in the 10\,ms to 1\,s range as the system's planning layer. xApps in this loop perform deliberative reasoning rather than immediate reaction. These xApps correlate observations across time, anticipate demand shifts, resolve conflicts among competing reflexive actions, and adjust parameters based on a coherent, evolving view of the network. To support this reasoning, working memory in this tier is organized into two levels. Fast memory holds the active decision context, such as the current beam configuration and scheduler state, while pooled memory retains broader context from recent control intervals. Through the unified memory fabric, xApps can access a bounded window of sensory memory maintained by the reflexive loop and selected representations promoted to working memory. Rather than relying only on pre-aggregated metrics, xApps combine these observations with higher-layer context such as user trajectories and slice-level Service Level Agreement (SLA) status, supporting spatiotemporal reasoning that moves beyond reactive control.

\subsection{Evolutionary Loop}
\label{subsec:evolutionary-loop}

Encompassing the entire hierarchy is the evolutionary loop, operating across minutes, hours, or days as the system's learning center. Over these extended timescales, wireless environments exhibit long-horizon dynamics, from recurring traffic patterns to gradual channel drift, that short-term memory cannot retain. The evolutionary loop therefore maintains long-term memory in three complementary forms. Episodic memory records what the network experienced, storing significant interaction logs and anomaly traces as vector embeddings in data lakes. Semantic memory encodes what the network knows, representing structural facts such as topology, base station capabilities, and 3rd Generation Partnership Project (3GPP) protocol constraints in knowledge graphs. Procedural memory captures what the network has learned to do, holding trained policy networks and model weights that embody accumulated behavioral strategies. Non-real-time RAN applications (rApps), the intelligence units native to this loop, operate offline to query episodic and semantic memory, identify long-term performance gaps, and retrain or fine-tune procedural models accordingly. Updated policies then propagate down to the contextual and reflexive loops, enabling the system's reflexes and reasoning to adapt based on accumulated experience.

\subsection{Dual-Pathway Coordination}
\label{subsec:cross-loop-coordination}

\begin{figure}[t]
\centering
\includegraphics[width=0.9\columnwidth]{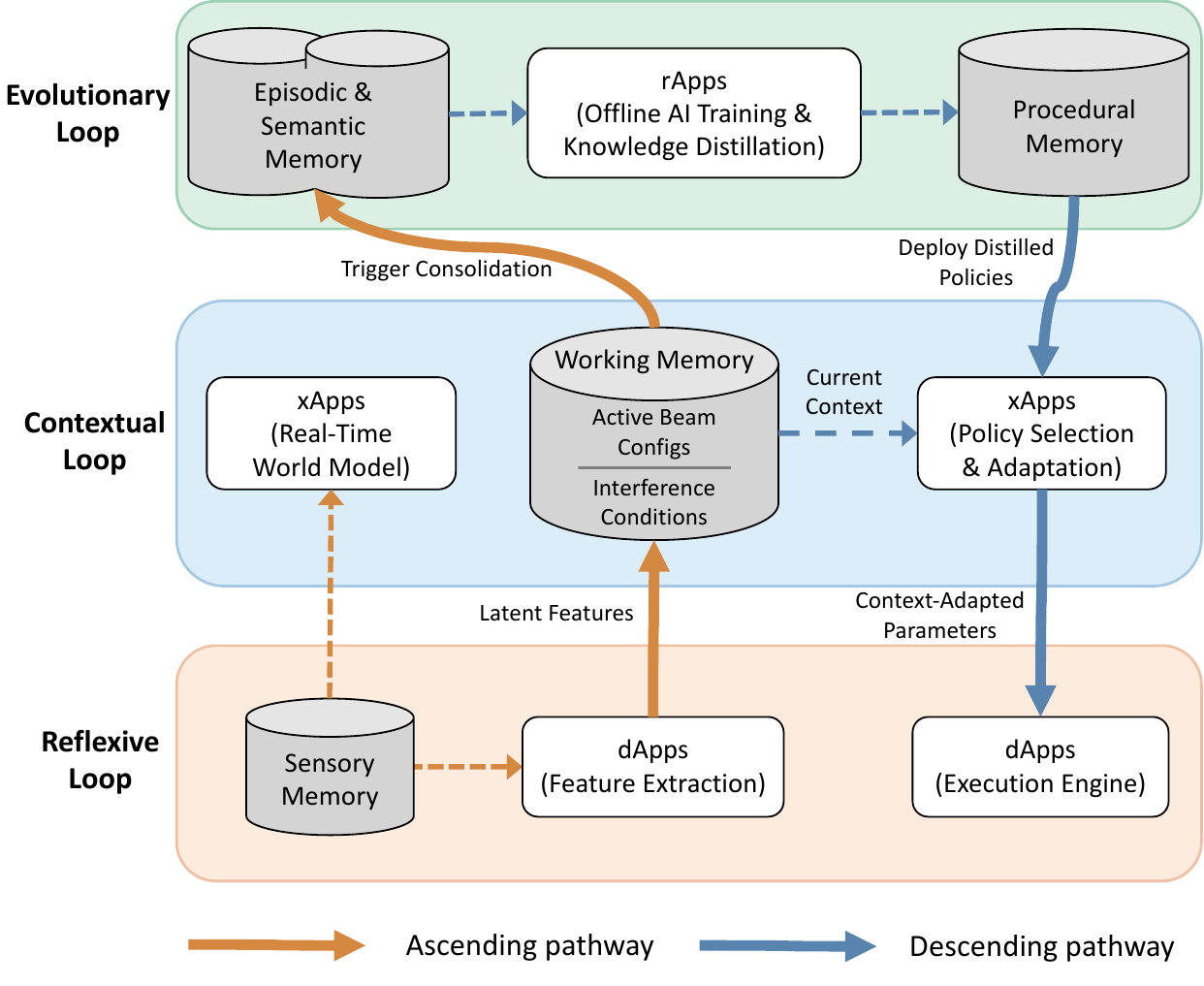}
\caption{Dual-pathway coordination mechanism connecting the reflexive, contextual, and evolutionary loops. }
\label{fig:pathway}
\end{figure}

The power of this architecture lies in the flow of memory that connects the three multi-scale cognitive loops, transforming raw data into actionable intelligence through a dual-pathway coordination mechanism. This dual flow ensures a consistent view of reality across the real-time, near-real-time, and long-term horizons, as shown in Fig.~\ref{fig:pathway}.

The ascending pathway facilitates perception and consolidation. It begins in the reflexive loop, where dApps extract latent features such as Doppler spread and multipath signatures from raw I/Q samples in sensory memory. Unlike the fixed interface abstractions discussed in Section~\ref{sec:cognitive-gap}, this extraction does not replace access to the underlying raw data. Through the unified memory fabric, the contextual loop receives selected features together with a bounded window of recent sensory memory, preserving relevant spatial and temporal context without requiring all raw samples to be serialized into interface messages. This allows xApps to construct a real-time world model from selected radio observations and higher-layer context. Consolidation from working memory to long-term memory is triggered selectively by episode significance. The contextual loop evaluates and retains selected episode representations asynchronously for future retrieval, allowing the reflexive loop to continue local processing without waiting for cross-tier persistence. This selective process avoids flooding long-term memory with routine observations while ensuring that novel events are captured for future learning.

The descending pathway governs control and skill transfer. The evolutionary loop trains AI policies offline using historical data from episodic and semantic memory. The resulting full-scale models, optimized for accuracy rather than latency, are distilled into compact variants through knowledge distillation before being stored in procedural memory. The contextual loop reads the distilled policies from procedural memory and, using current working memory context such as active beam configurations and interference conditions, selects and adapts the appropriate policy variant for the current network state. The resulting context-adapted parameters are then written to a shared memory region from which dApps in the reflexive loop load them for execution.

\begin{table*}[t]
\centering
\caption{Mapping of Cognitive Loops to Memory Types, Hardware, and Access Patterns}
\label{tab:cognitive-mapping}
\resizebox{\textwidth}{!}{%
\begin{tabular}{lllll}
\toprule
Cognitive Loop & Time Horizon & Memory Type & Physical Implementation & Access Pattern \\
\midrule
\multirow{2}{*}{Reflexive Loop} & \multirow{2}{*}{Below 10~ms (real-time)} & \multirow{2}{*}{Sensory} & \multirow{2}{*}{FPGA BRAM/URAM + CXL Type-2 device memory} & Read from Sensory \\
             &              &         &                               & Write to Sensory \\
\midrule
\multirow{2}{*}{Contextual Loop} & \multirow{2}{*}{10~ms--1~s (near-real-time)} & \multirow{2}{*}{Working} & \multirow{2}{*}{GPU HBM + CXL Memory Pool} & Read from Sensory \\
               &              &         &                               & Read/Write to Working \\
\midrule
\multirow{2}{*}{Evolutionary Loop} & \multirow{2}{*}{Minutes/Hours/Days (non-real-time)} & \multirow{2}{*}{Long-Term} & \multirow{2}{*}{CPU DRAM + NVMe storage} & Read from Episodic/Semantic \\
                  &              &         &                               & Write to Procedural \\
\bottomrule
\end{tabular}%
}
\end{table*}

\section{Unified Memory Fabric}
\label{sec:unified-memory}

In this section, we elaborate on the unified memory fabric designed for the proposed multi-scale cognitive hierarchy, which encompasses sensory memory, working memory, and long-term memory. Table~\ref{tab:cognitive-mapping} summarizes the mapping of cognitive loops to memory types, physical implementations, and access patterns.

\subsection{Sensory Memory}
\label{subsec:sensory-memory}

In AI-RAN, massive streams of I/Q samples arrive from the Radio Unit (RU), and copying this data through multiple memory hops introduces latency that strains real-time processing deadlines. Sensory memory addresses this by serving as a bounded, high-fidelity circular buffer that is continuously overwritten as new samples arrive. Before entries are overwritten, xApps in the contextual loop can inspect the retained buffer window and trigger feature extraction, promoting selected representations to working memory. This design is realized as a two-level physical hierarchy. The first level uses on-chip FPGA Block RAM (BRAM) and UltraRAM (URAM) as the immediate capture buffer, where hard reflex dApps perform initial signal processing directly on the stream. The second level uses CXL Type-2 device memory, which maps the FPGA device memory into a shared address space. This mapping provides load/store access to the current buffer window for compatible soft-reflex dApps and xApps without an intermediate copy.

\subsection{Working Memory}
\label{subsec:working-memory}

Working memory is the active workspace where perception meets reasoning. We structure it as a two-tier system, with fast memory backed by GPU high-bandwidth memory (HBM) and pooled memory backed by CXL-attached capacity. Fast memory holds the volatile state required for active inference, retaining the immediate decision context including current beam configurations, active scheduler state, and real-time channel estimates. For xApps that incorporate LLM-based reasoning, it also maintains the attention KV cache needed for low-latency token generation. Pooled memory provides elastic context that extends beyond local HBM limits, retaining broader situational awareness such as recent mobility trajectories, slice SLA status, and global interference maps accumulated over recent control intervals. This organization allows xApps to combine selected observations from the sensory-memory window with recently accumulated context, supporting spatiotemporal reasoning across control intervals.

\subsection{Long-Term Memory}
\label{subsec:long-term-memory}

While working memory retains the context of recent control intervals, long-term memory provides the persistent foundation for continuous learning and inference. We organize long-term memory into three complementary forms that together capture what the network experienced, what it knows, and what it has learned to do. Episodic memory retains selected observations and anomaly traces from significant interactions, together with the context, action, and outcome associated with each episode. Stored in vector databases, these records support Retrieval-Augmented Generation (RAG), allowing rApps to compare a current situation with prior episodes, inform subsequent decisions, and identify recurring failure modes. Semantic memory stores structured, explicit knowledge, implemented as knowledge graphs. It retains static or slowly changing facts, such as network topology, base station capabilities, and 3GPP protocol specifications. The rApps use this repository to execute symbolic reasoning, such as ascertaining if a candidate policy action violates a specific compliance mandate before execution. Procedural memory serves as the repository for the network's learned capabilities, embodied as AI model weights and policy networks. These models are written by rApps after offline training and read by the contextual loop during policy selection. Physically, these three forms reside on high-capacity non-volatile memory express (NVMe) storage for persistence, with CPU dynamic random-access memory (DRAM) serving as an active-access tier that caches frequently queried structures, such as vector database indices, knowledge graph partitions, and frequently used policy variants.

\subsection{Memory Interconnects}
\label{subsec:unified-memory-interconnects}

\begin{figure}[t]
\centering
\includegraphics[width=0.99\columnwidth]{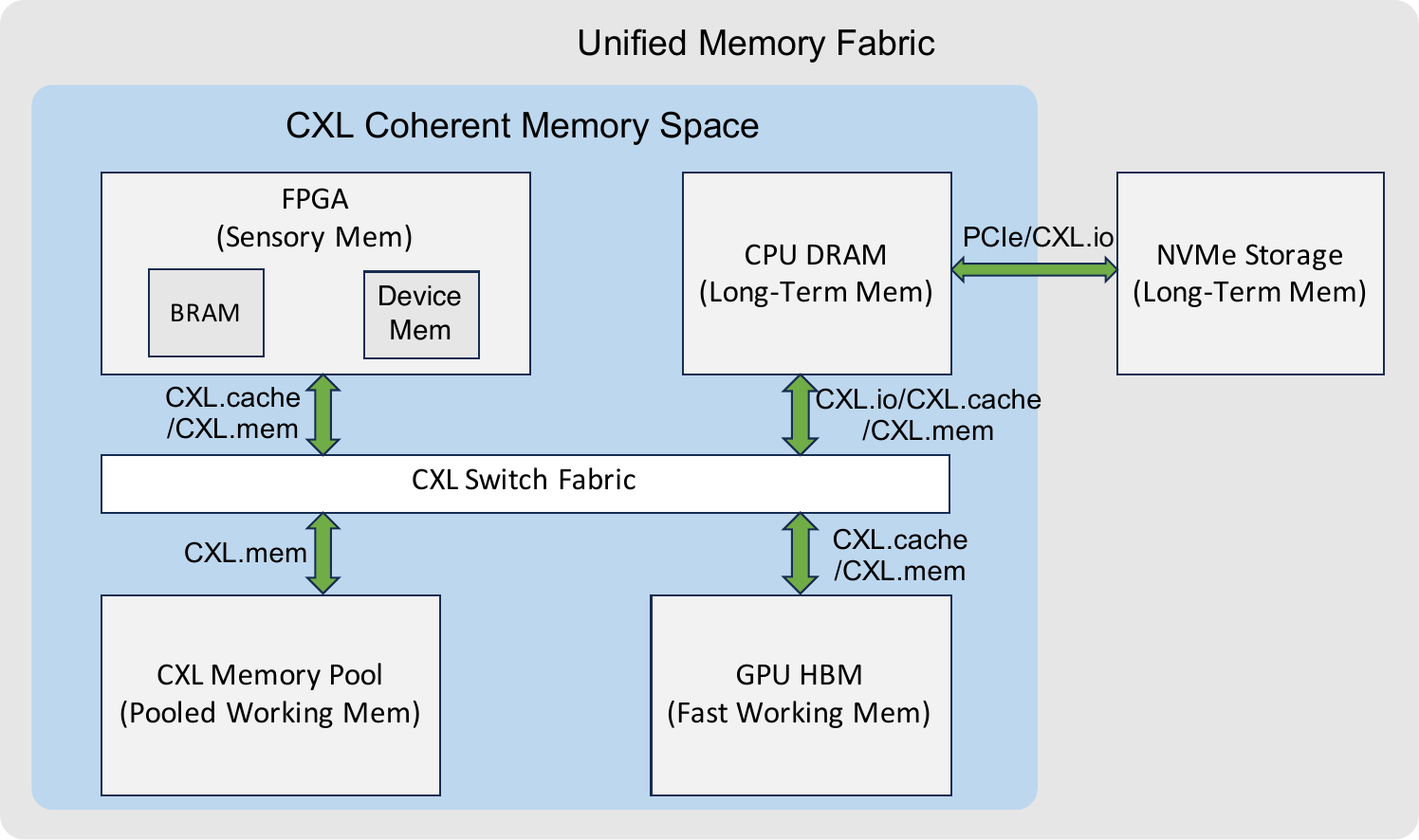}
\caption{An illustration of a CXL-based memory fabric, where a CXL switch connects FPGA, GPU, and CPU to support coherent access to selected memory across compatible devices.}
\label{fig:cxl-architecture}
\end{figure}

The three memory tiers described above require an interconnect that can provide coherent access to selected state across compatible heterogeneous devices, with CXL serving as one such candidate. As illustrated in Fig.~\ref{fig:cxl-architecture}, a CXL switch connects FPGA, GPU, and CPU through the CXL protocols. CXL.cache supports coherent access to host memory from a device, CXL.mem supports host access to device-attached memory, and CXL.io handles discovery and configuration.

This fabric enables bidirectional data flow across the cognitive hierarchy. On the ascending path, the FPGA's device memory is mapped into a coherent address region accessible to compatible devices via CXL Type-2 attach mode, allowing compatible GPU and CPU domains to access the current sensory buffer window through the CXL memory path. The contextual loop correlates these observations with recent context held in working memory, where CXL memory pooling provides elastic capacity beyond local GPU HBM. When an episode-significance score exceeds a configured threshold, the consolidation mechanism schedules selected episodes for asynchronous persistence. A lower-priority operation then migrates them from pooled working memory to NVMe-backed long-term storage via PCIe/CXL.io. The migration can be batched or deferred when computational or interconnect resources are heavily utilized. On the descending path, updated policies from the evolutionary loop are read from long-term memory, adapted by the contextual loop, and written to a shared memory region from which dApps in the reflexive loop load them for low-latency execution. These transfers can use shared-memory access for selected state, reducing reliance on serialized message updates in the real-time path.

% This unified interconnect architecture enables a paradigm shift from ``Networked Disaggregation'' to ``Hyper-converged Integration.'' While the O-RAN Alliance has standardized the \emph{logical} separation of control (RIC) and data planes (DU), the \emph{physical} distribution across servers introduces a latency tax. By replacing network sockets with CXL memory buses, we collapse the physical distance between FPGA, GPU, and CPU to zero while keeping their functional roles distinct.

\section{Case Study}
\label{sec:case-study}

Returning to the stadium scenario introduced in Section~\ref{sec:cognitive-gap}, we evaluate the benefit of the proposed memory hierarchy through a proof-of-concept study that treats CXL transport as transparent and focuses on cognitive-level gains. Finite CXL bandwidth, PCIe lane availability, and transport contention are not modeled in this proof-of-concept, and hence the resulting spectral-efficiency gains represent an upper-bound estimate of cognitive-level benefits rather than end-to-end deployment performance. We consider that a 64-antenna next-generation NodeB (gNB) forms eight analog beams over 100 subcarriers, yielding a beam-SNR matrix with 800 elements per snapshot. We compare two designs under the same objective of maximizing spectral efficiency. An interface-based agent receives only compressed indicators (e.g., average SINR), while a memory-based agent reads the full beam-SNR matrix through unified memory. The spatial experiment evaluates whether richer observability from sensory memory improves interference diagnosis and immediate remediation, and the temporal experiment evaluates whether episodic retrieval from long-term memory improves adaptation across recurring weekly events.

\begin{figure}[t]
\centering
\includegraphics[width=0.90\columnwidth]{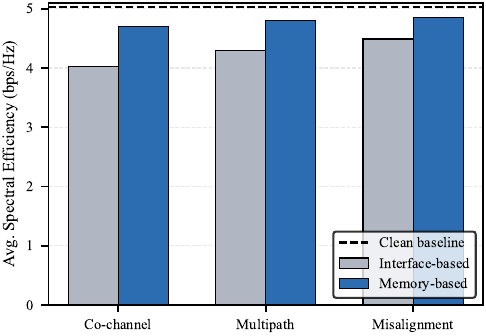}
\caption{Average spectral efficiency under three interference regimes. 
%The interface-based agent observes only compressed indicators and applies uniform remediation, while the memory-based agent reads the full beam-SNR matrix and applies targeted mask-based remediation. Full-state observability provides consistent gains across co-channel interference, multipath fading, and beam misalignment.
}
\label{fig:spatial-throughput}
\end{figure}

In the spatial experiment, we inject three interference types that produce distinct patterns in the raw beam-SNR matrix, which resides in sensory memory. Co-channel interference degrades two to three adjacent beams across nearly all subcarriers. Multipath fading degrades clustered subcarrier bands across all beams. Beam misalignment causes a deep collapse in one dominant beam. These patterns are visually separable in the matrix domain but become ambiguous after compression to average SINR. The interface-based agent therefore applies a uniform power boost regardless of root cause, recovering only part of the lost capacity. The memory-based agent identifies each interference type by inspecting the beam-subcarrier pattern in the full matrix and applies targeted mask-based recovery, selectively compensating the affected elements for each case. As shown in Fig.~\ref{fig:spatial-throughput}, full-state observability yields spectral efficiency gains of 8\% to 17\% across the three interference regimes. The relative gain scales with the fraction of beam-subcarrier elements affected. Co-channel interference degrades three adjacent beams across all subcarriers, producing the largest gain, whereas beam misalignment concentrates degradation in a single beam and yields the smallest percentage improvement despite the deepest per-element recovery. %The same trials reveal a large separability gap in interference diagnosis, where the memory-based classifier reaches 100.0\% accuracy while the interface-based classifier remains near chance at 33.6\%.

\begin{figure}[t]
\centering
\includegraphics[width=0.9\columnwidth]{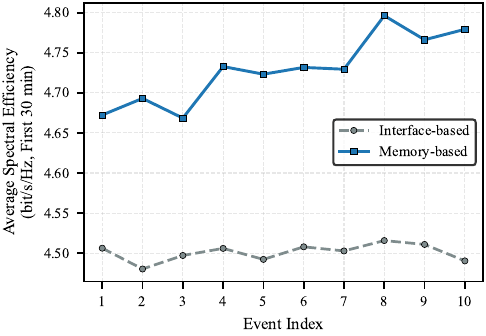}
\caption{Average spectral efficiency over the first 30 minutes of ten recurring events. 
%The interface-based agent repeatedly cold-starts from compressed observations, while the memory-based agent warm-starts from episodic retrieval. Persistent memory yields faster adaptation and better early-event throughput.
}
\label{fig:temporal-convergence}
\end{figure}

In the temporal experiment, we model ten recurring weekly events, each spanning 180 one-minute control intervals that follow the same crowd-mobility stage sequence, moving through pre-game arrival, game-time congestion, halftime redistribution, second-half surge, and post-game departure. The stage sequence is identical across events, but each event is assigned a distinct beam-subcarrier interference pattern that determines which beams and subcarrier clusters are affected. The interface-based agent is memoryless at event start and must re-learn effective control from default settings in each episode. The memory-based agent queries episodic memory in long-term memory to retrieve the most similar past event, measured by cosine distance between the observed beam-SNR deviation profile and stored episode profiles, and initializes policy parameters from that retrieved experience. Fig.~\ref{fig:temporal-convergence} shows the average spectral efficiency over the first 30 minutes of each event, where adaptation quality has the strongest operational impact. The interface-based curve remains nearly flat because each event repeats cold-start behavior. In contrast, the memory-based curve rises over early events and then saturates once the episodic store covers the range of encountered interference conditions. It consistently achieves higher average spectral efficiency during this early-event period. This pattern indicates that retrieved historical context reduces convergence time and improves early-stage performance.

Together, these results confirm that richer spatial observability enables cause-aware control and persistent temporal context accelerates adaptation across recurring events, supporting the case for unified memory as an architectural prerequisite for multi-scale cognition in agentic AI-RAN.

\section{Open Issues}
\label{sec:open-issues}

While the unified memory architecture addresses the cognitive gap in agentic AI-RAN, realizing this vision involves overcoming significant challenges. The convergence of heterogeneous computing and high-dimensional AI reasoning introduces new trade-offs in coherency management, model synchronization, multi-tenancy isolation, security, and cross-node cognitive continuity that together define the open research agenda for 6G agentic networks.

\subsection{The Challenge of Coherency Tax}

Coherent memory sharing across CPU, GPU, and FPGA is fundamental to our architecture, but maintaining cache coherency across these distinct silicon dies incurs a coherency tax, namely the bandwidth and latency overhead generated by snooping protocols. The practical latency and determinism of a CXL-based path also depend on endpoint support, topology, firmware, and traffic contention. For example, in massive multiple-input multiple-output (MIMO) scenarios, if the reflexive loop updates channel estimates too frequently, coherency traffic can flood the interconnect, starving the contextual loop and paradoxically increasing latency. Future interconnect standards may need to expose programmable consistency models, trading slight staleness for bus efficiency. Depending on use, state for immediate scheduling and control must remain current, whereas mobility trajectories, load trends, interference summaries, and historical episode features may tolerate bounded staleness. Topology and hardware capabilities require less frequent refresh. Defining staleness bounds that each cognitive tier can tolerate without compromising real-time RAN correctness remains an open question.

\subsection{Model Synchronization Across Tiers}

Our three-tier cognitive architecture introduces a model synchronization challenge that traditional RAN systems never faced. Model updates propagate downward through the cognitive hierarchy, from the evolutionary loop through the contextual loop to the reflexive loop, with each tier adapting and executing the received policies at its own timescale. This propagation is inherently non-atomic, as each stage requires processing time while every tier continues serving its real-time workload. During the transition, adjacent tiers inevitably operate on different model generations, yet no tier can pause to wait for the others to converge. How to orchestrate this cascading synchronization, including when to trigger each stage, in what order tiers should switch, and how to maintain service continuity while the update propagates, remains an open problem without established methodology in real-time AI systems.

\subsection{Multi-Tenancy in Unified Memory}

Within the unified memory fabric, AI workloads and RAN processing become co-tenants of the same physical memory. This co-location creates noisy neighbor effects, where one workload's memory activity degrades the performance of another. The two sides exhibit fundamentally different access patterns. RAN physical layer processing demands deterministic bursts aligned with radio-processing deadlines, whereas AI workloads, ranging from real-time inference to offline model training, generate irregular and often sustained memory traffic. When these patterns collide, a model retraining job can saturate CXL bandwidth and evict cache lines needed for real-time beamforming, causing AI workloads to degrade RAN performance. Designing memory-aware arbitration that guarantees bounded latency for real-time RAN functions while allowing AI workloads to consume remaining bandwidth elastically remains an open problem.

\subsection{Security in Borderless Memory}

Unified memory replaces the process-level isolation that traditionally separates RAN workloads with a shared address space accessible to all co-located agents. This architectural choice expands the attack surface. A compromised xApp with access to shared working memory could perform a buffer over-read to extract sensitive user embeddings, or inject adversarial noise into the beamforming weights of another slice. Addressing this threat requires extending confidential computing to heterogeneous memory pools, bringing trusted execution environments beyond the CPU to cover GPU HBM and CXL-attached modules, so that agents can reason about sensitive data within encrypted enclaves without exposing it to the host operating system or co-located workloads.

\subsection{Cognitive Continuity Across Node Boundaries}

The unified memory fabric operates within a single physical node. However, at network scale, multiple DUs, CUs, and RICs must still coordinate across physical boundaries. At the node boundary, the shared-memory paradigm necessarily yields to message passing, and agents lose the direct observability that defines intra-node cognition. Within a node, agents perceive the full physical-layer context through shared memory, but across node boundaries they must fall back on compressed representations exchanged through interfaces such as E2 and A1. This reintroduces the very semantic bottleneck that unified memory eliminates internally. Extending cognitive continuity beyond a single node remains a fundamental challenge for scaling this architecture to network-wide intelligence. One direction is a distributed shared-memory abstraction that provides an access model in which selected states are addressable across node boundaries under explicit ownership and update rules. Another direction is a semantic-preserving inter-node protocol that exchanges selected states, such as features, episode representations, and policy updates, together with their scope, freshness, and provenance, allowing receiving agents to interpret the exchanged states and combine them with local observations across node boundaries.

%\subsection{Research Roadmap}
%
%The realization of this vision aligns with the CXL technology roadmap. In the near term, CXL 3.0 (available in commercial silicon today) already supports memory pooling and multi-headed device sharing, enabling the basic sensory-to-working memory path described in this article. In the medium term, CXL 3.1 introduces enhanced fabric management and per-device QoS controls that would address the multi-tenancy challenges discussed above. Looking further ahead, CXL 4.0 and beyond are expected to deliver full fabric-level coherency with lower latency, which would enable the tight cross-tier coordination required for the most demanding reflexive loop scenarios. Each generation brings the vision closer to practical deployment without requiring a disruptive architectural change.

\section{Conclusion}
\label{sec:conclusion}

The transition from 5G to 6G marks the evolution of the RAN from a passive transmission medium into an intelligent entity capable of autonomous action. This article identified the cognitive gap inherent in disaggregated architectures, where lack of persistent memory and state isolation between sensing and reasoning inflict temporal amnesia and spatial blindness on agents in AI-RAN. We proposed a unified memory architecture that dissolves this dichotomy by mapping cognitive memory hierarchies onto heterogeneous computing fabric interconnected via CXL. By supporting selected cross-tier state sharing, this architecture gives agents access to relevant physical context and preserves experience across decision cycles. Realizing this vision requires addressing open challenges in coherency management, model synchronization, multi-tenant isolation, and confidential computing. Success in these areas will transform the RAN into a truly autonomous cognitive system for the 6G era. %Future work will focus on hardware prototyping with CXL-enabled platforms and system-level benchmarking to validate the performance gains of memory-centric coordination across real-world deployment scenarios.

\bibliographystyle{IEEEtran}

\vspace{-10mm}

\begin{IEEEbiographynophoto}
{Xijun Wang} (wangxijun@mail.sysu.edu.cn) received the B.S. degree from Xidian University in 2005 and the Ph.D. degree from Tsinghua University in 2012. He is an Associate Professor with Sun Yat-sen University, China.
\end{IEEEbiographynophoto}

\vspace{-10mm}

\begin{IEEEbiographynophoto} 
{Zhaoyang Liu} (liuzhy86@mail2.sysu.edu.cn) received the B.E. degree from Sun Yat-sen University in 2022, where he is pursuing the Ph.D. degree.
\end{IEEEbiographynophoto}

\vspace{-10mm}

\begin{IEEEbiographynophoto}
{Chenyuan Feng} (c.feng@exeter.ac.uk) received the Ph.D. degree from Singapore University of Technology and Design in 2021. She is a researcher with the University of Exeter, U.K.
\end{IEEEbiographynophoto}

\vspace{-10mm}

\begin{IEEEbiographynophoto}
{Xiang Chen} (chenxiang@mail.sysu.edu.cn) received the B.E. and Ph.D. degrees from Tsinghua University in 2002 and 2008, respectively. He is a Full Professor with Sun Yat-sen University, China.
\end{IEEEbiographynophoto}

\vspace{-10mm}

\begin{IEEEbiographynophoto}
{Howard H. Yang} (haoyang@intl.zju.edu.cn) received the Ph.D. degree from Singapore University of Technology and Design in 2017. He is an Assistant Professor with Zhejiang University, China.
\end{IEEEbiographynophoto}

\vspace{-10mm}

\begin{IEEEbiographynophoto}
{Tony Q. S. Quek} (tonyquek@sutd.edu.sg) received the Ph.D. degree from the Massachusetts Institute of Technology in 2008. He is the Cheng Tsang Man Chair Professor with Singapore University of Technology and Design, Singapore.
\end{IEEEbiographynophoto}

\end{document}